%% file: arxivv3.tex
\UseRawInputEncoding 
\documentclass[%
superscriptaddress,  aps, prapp,twocolumn,longbibliography
]{revtex4-1}

\usepackage{graphicx, upgreek}
\usepackage{dcolumn}
\usepackage{bm}
\usepackage{siunitx}
\usepackage{color}
\usepackage{multirow}
\usepackage{makecell}

\input{defs_thesis}

\begin{document}

\preprint{AIP/123-QED}

\title
{Electron spin resonance of P donors in isotopically purified Si detected by contactless photoconductivity}

\author{ Philipp Ross}
\affiliation{London Centre for Nanotechnology, 17-19 Gordon St., London WC1H 0AH }
\author{Brendon C. Rose}
\affiliation{Department of Electrical Engineering, Princeton University, Princeton, New Jersey 08544, USA}

\author{Cheuk C. Lo}
\affiliation{London Centre for Nanotechnology, 17-19 Gordon St., London WC1H 0AH }
\affiliation{Department of Electronic and Electrical Engineering, University College London, Torrington Place, London WC1E 7JE}
\author{Mike L. W. Thewalt}
\affiliation{Simon Fraser University, Burnaby, British Columbia V5A 1S6, Canada}	
\author{Alexei M. Tyryshkin}
\affiliation{Department of Electrical Engineering, Princeton University, Princeton, New Jersey 08544, USA}	
\author{Stephen A. Lyon}
\affiliation{Department of Electrical Engineering, Princeton University, Princeton, New Jersey 08544, USA}	
\author{John J. L. Morton\thanks{Address correspondence to jjl.morton@ucl.ac.uk}}
\affiliation{London Centre for Nanotechnology, 17-19 Gordon St., London WC1H 0AH }
\affiliation{Department of Electronic and Electrical Engineering, University College London, Torrington Place, London WC1E 7JE}

\date{\today}

\begin{abstract}

Coherence times of electron spins bound to phosphorus donors have been measured, using a standard Hahn echo technique, to be up to 20~ms in isotopically pure silicon with [P]$ = 10^{14}$~cm$^{-3}$ and at temperatures $\leq 4$~K. Although such times are exceptionally long for electron spins in the solid state, they are nevertheless limited by donor electron spin-spin interactions. Suppressing such interactions requires even lower donor concentrations, which lie below the detection limit for typical electron spin resonance (ESR) spectrometers. Here we describe an alternative method for phosphorus donor ESR detection, exploiting the spin-to-charge conversion provided by the optical donor bound exciton transition. We characterize the method and its dependence on laser power and use it to measure a coherence time of $T_2 = \SI{130}{ms}$ for one of the purest silicon samples grown to-date ([P]$ = \SI{5e11}{cm^{-3}}$). We then benchmark this result using an alternative application of the donor bound exciton transition: optically polarising the donor spins before using conventional ESR detection at \SI{1.7}{K} for a sample with  [P]$ = \SI{4e12}{cm^{-3}}$, and measuring in this case a $T_2$ of \SI{350}{ms}. In both cases, $T_2$ is obtained after accounting for the effects of magnetic field noise, and the use of more stable (e.g. permanent) magnets could yield even longer coherence times.
\end{abstract}

\pacs{Valid PACS appear here}
\maketitle

\section{Introduction} 
A number of factors are critical in the measurement of long coherence times in solid state spin systems, including instrumental challenges such as stability in the magnetic field and microwave phase, as well as host crystal purity. In the case of silicon, isotopically enriched \sitwoeight\ crystals~\cite{Becker2010} have been used to extend the coherence time limits of both nuclear and electron donor spins in bulk samples \cite{Steger2012, Saeedi2013, Tyryshkin2012, Wolfowicz2013}, as well as in nanoscale single donor devices \cite{Muhonen2014}. However, while purifying the host environment is important, spin-spin interactions between same-species donors also play a limiting factor, due to the finite donor concentration within the sample. 
For example, from a \sitwoeight\ sample with [P]$ = 10^{14}$\,cm$^{-3}$, a coherence times of up to \SI{\sim 20}{ms} was measured~\cite{Tyryshkin2012}, and shown to be limited by dipolar interactions between phosphorus donor spins. These interactions cannot be reversed in a standard Hahn echo measurement (this effect is also known as `instantaneous diffusion'~\cite{Kurshev1992,Schweiger2001}), however, their effect can be reduced by artificially reducing the spin-concentration of the sample using a spin echo sequence with shortened refocussing pulse. Such methods enable an estimate for the expected coherence times in more dilute samples, and led to inferred decay times of approximately 1 second~\cite{Tyryshkin2012}. However, the method does not account for other effects which may be present in samples with low donor centrations, such as donor-acceptor recombination. 

Conventional electron spin resonance (ESR) spectrometers are close to their detection limits for spin concentrations in the range of $10^{12}$ cm$^{-3}$ and above, and thus new detection methods are needed in order to directly measure spin donor concentration times in samples with lower doping densities. Electrically detected magnetic resonance (EDMR) has been shown as a technique to study small numbers of donor electron spins~\cite{Stegner2006} down to the level of 100 donors~\cite{McCamey2006}, however, most experiments have relied on coupling to spin-active defects at the Si/SiO$_2$ interface for readout, and in such cases \ttwo\ is typically of order 1~$\mu$s~\cite{Paik2010}. EDMR methods have recently been combined with the use of donor bound exciton \DX{} spectroscopy to measure intrinsic donor spin coherence times~\cite{Lo2015}, leading to a maximum \ttwo\ of about 1.5~ms, in that case still limited by the donor concentration.


In this Letter, we report the coherence time measurement of two \textsuperscript{28}Si samples with [P] = \SI{5e11}{cm^{-3}} and  \SI{4e12}{cm^{-3}}, using spin selective ionisation via the donor bound exciton transition \DX{} to optically polarize the spin ensemble beyond the thermal equilibrium value ~\cite{Yang2008,Lo2015}. We then perform either a conventional ESR experiment (for the higher concentration sample) or measure the donor electron spin state via the spin-dependent photoconductivity following \DX{} excitation~\cite{Steger2012,Lo2015,Franke2016}, using a contactless technique where the sample is inserted into a parallel plate capacitor. We fully characterise this contactless measurement method, studying the on- and off-resonance conductivity of the sample as a function of laser power. We perform Hahn echo coherence time measurements and find a coherence time of  $T_2 = \SI{130}{ms}$ for the lower doped sample (using full ($\pi$) refocussing pulses) at \SI{4.5}{K}. 
We compare this time with that measured from the higher doped sample using hyperpolarized ESR, where we find a $T_2$ of \SI{350}{ms}. To our knowledge, these are the longest coherence times reported to date for an electron spin in a solid state away from a clock transition \cite{Wolfowicz2014}, and using full refocussing pulses \cite{Tyryshkin2012}.

\section{Contactless photoconductive \DX{} detection}

\begin{figure}[t]
\includegraphics{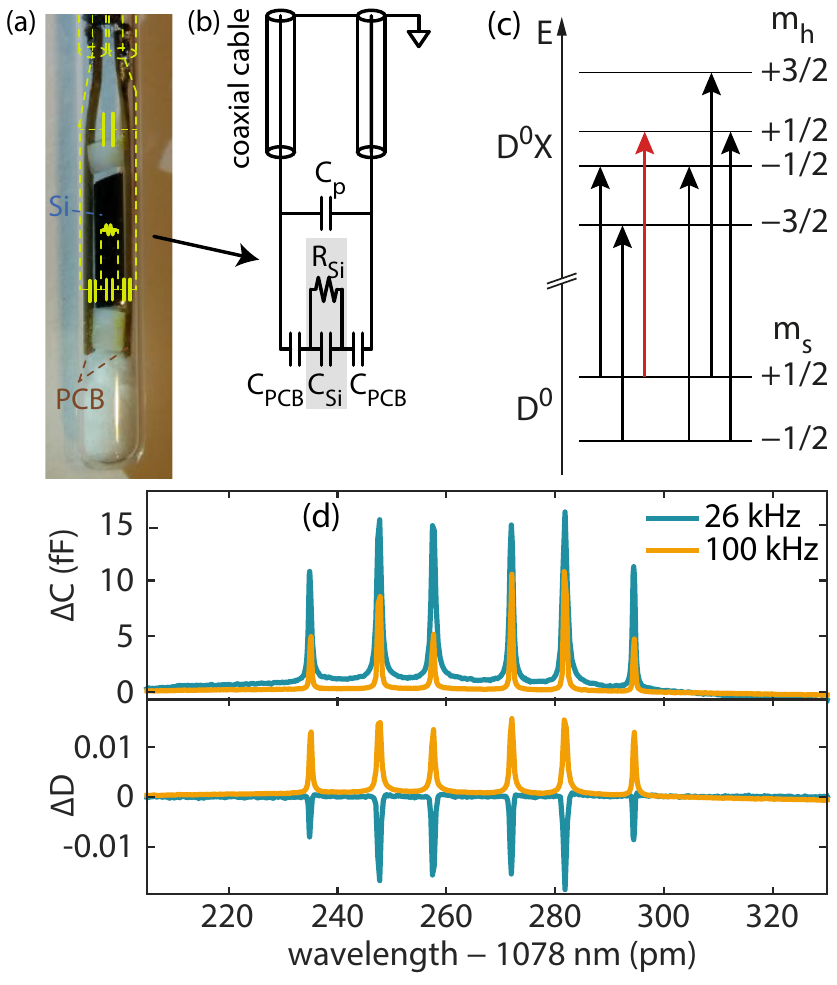}%
\caption{\label{fig:Setup} The experimental setup and \DX{} spectrum measured using contactless photoconductivity readout. (a) Photograph of the bottom of the probe stick and (b) its corresponding circuit diagram. (c) The six allowed \DX{} transitions in a magnetic field, sorted by transition energy. The transition highlighted in red is used for ESR detection. 
(d) Changes of the capacitance $C$ and loss tangent $D$  for a \textsuperscript{28}Si:P ([P]~$=\SI{2e14}{cm^{-3}}$) in a magnetic field of \SI{340}{mT} and a laser intensity of about \SI{40}{\upmu W/mm^2}.  
}
\end{figure}

The basis for the hyperpolarization and the spin-to-charge conversion used here is the donor bound exciton transition. Neutral donors in silicon can be optically excited to the bound exciton (\DX{}) state  in which two electrons and a hole bind to the donor. In the (\DX{}) ground state the two electrons form a spin singlet and the hole spin ($J=3/2$) determines the Zeeman splitting in an external magnetic field, as depicted in Fig.~\ref{fig:Setup}(c). The long bound exciton lifetime (\SI{272}{ns}~\cite{Schmid1977}) and relatively small inhomogeneous broadening result in \DX{} optical transitions that are sufficiently narrow to enable the excitation of the donor selectively on its electron spin state \cite{Lo2015} (and in low-strain \sitwoeight, even the donor nuclear spin state can be resolved \cite{Steger2012, Saeedi2013}). The \DX{} recombines via an Auger process, ejecting an electron into the conduction band and leaving behind the ionized donor, producing a change in sample conductivity.

We capacitively measure the conductivity change on \DX{} resonance using the setup shown in Fig.~\ref{fig:Setup}(a): The silicon crystal ($\SI{2}{mm}\times\SI{2}{mm}\times\SI{10}{mm}$) is mounted between two PCBs separated by two teflon spacers to minimize any applied stress to the sample, all placed within a quartz tube of \SI{5}{mm} diameter. The copper electrodes of the PCB face outward to avoid direct electrical contact with the sample and are contacted via two cryogenic stainless steel coaxial cables. The probe stick is inserted into a dielectric ring microwave resonator and cooled to \SI{4.5}{K}.  We measure the capacitance $C$ and loss tangent $D = \mathrm{Re}(Z)/|\mathrm{Im}(Z)|$ of the sample capacitor with an Agilent E4980D LCR meter ($V_\mathrm{AC}=\SI{1}{V}$, $V_\mathrm{DC}=\SI{0}{V}$, no resulting avalanche carrier generation), at some modulation frequency, $f_m$, and excite the \DX{} transition using a NKT Boostik fibre laser with a nominal linewidth of \SI{70}{kHz}. 

\begin{figure}[t]
\includegraphics{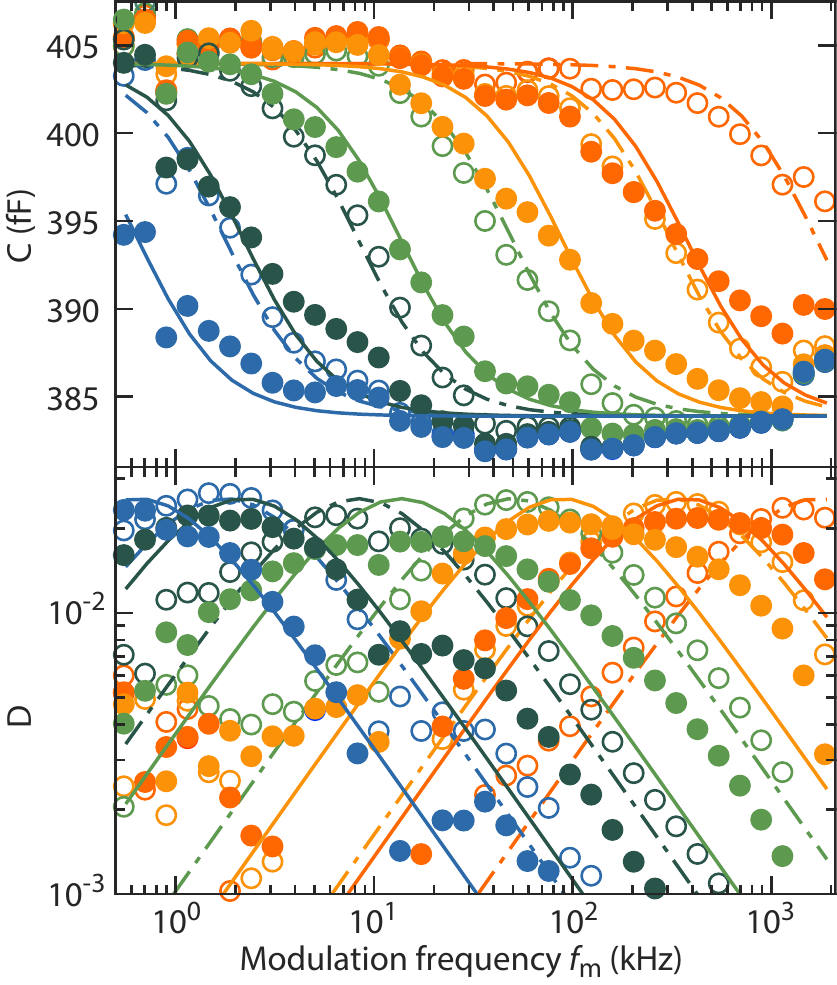}%
\caption{\label{fig:freqSweep} The  total capacitance, $C$, and loss tangent, $D$ measured as a function of probe frequency and laser power, both on-resonance (empty circles) and off-resonance (filled circles) with the \DX{} transition at $B=\SI{0}{T}$.  The laser power increases (non lineary, see data points of Fig.\ \ref{fig:sigmaVpower}) from \SI{0.14}{\upmu W/mm^2} (blue) to \SI{3.5}{mW/mm^2} (red). $[\mathrm{P}] =\SI{3e15}{cm^{-3}}$.}
\end{figure}

Figure \ref{fig:Setup}(d) shows a typical \DX{} spectrum under an applied magnetic field. 
The six dipole-allowed \DX{} transitions are observable both as a change in capacitance and as a change of the loss tangent $D$. We find that depending on the modulation frequency, we either observe an increase or decrease of the loss tangent on-resonance, and explore this behaviour in more detail.
Figure \ref{fig:freqSweep} plots the capacitance and loss tangent of the sample as a function of modulaton frequency at $B=\SI{0}{mT}$, both on- and off-resonance with the \DX{} transition and for multiple different laser powers. We observe a step-like transition from a higher to a lower capacitance value with increasing frequency, coinciding with a peak in the loss tangent. This is a clear indication of a resonant phenomenon and we term this resonant frequency the `switching frequency', $f_\mathrm{s}$ in the following discussion. 
Increasing the applied laser power and bringing the laser on-resonance with the \DX{} transition are both characterised by an increase in $f_\mathrm{s}$. These observations explain the modulation frequency dependent behaviour of the \DX{} spectrum shown in Fig.\ \ref{fig:Setup}(d), i.e.\ that $C$ always increases on-resonance with the \DX{} transition, while $D$ either increases or decreases depending on whether the modulation frequency is larger or smaller than $f_\mathrm{s}$.

The underlying origin of the observed resonant behaviour can be traced back to a change of sample conductivity with laser power. We employ the circuit model shown in Fig.\ \ref{fig:Setup}(b) with the silicon sample modelled as a parallel circuit of some resistance $\mathrm{R_{Si}}$ and capacitance $C_\mathrm{{Si}}$, sandwiched between two PCB (of capacitance $C_\mathrm{{PCB}}$) and a parasitic capacitance $C_\mathrm{{p}}$ in parallel.  Under this model, we expect a larger measured capacitance for lower modulation frequencies because $\mathrm{R_{Si}}$ shorts the reactance associated with $C_\mathrm{{Si}}$ and thus the total capacitance is determined by the series connection of two $C_\mathrm{{PCB}}$. As the frequency increases, the reactance associated with $C_\mathrm{{Si}}$ becomes smaller, until the capacitive reactance dominates the sample impedance. Thus, for higher frequencies the total capacitance is lower since it is the series capacitance of $2\,C_\mathrm{{PCB}}$ and $C_\mathrm{{Si}}$. This transition occurs when the resistance $R_\mathrm{{Si}}$ and reactance $ \mathrm{X_{Si}} = 1/(\omega C_\mathrm{{Si}})$ are equal, leading to the switching frequency condition  $f_\mathrm{s} = \sigma_\mathrm{Si}/(2\pi\varepsilon_0\varepsilon_\mathrm{r})$, where $\sigma_\mathrm{Si}$ is the sample conductivity and  $\varepsilon_\mathrm{r} = 11.45$  is the dielectric constant for Si at \SI{4.5}{K}~\cite{krupka2006}, \footnote{The polarisability and concentration of phosphorus is negligible compared to the polarisability of the silicon lattice~\cite{Dhar1985}}. Using this  circuit model we can fit the whole data set (both $C(\omega)$ and $D(\omega)$ simultaneously) reasonably well with a single  value for $C_\mathrm{{p}}= \SI{290}{fF}$ and $C_\mathrm{{PCB}}= \SI{220}{fF}$. 
These values
 fit the expected parallel-plate capacitance by geometric considerations and secondly predict and match well the  measured reduction of probe stick capacitance after sample removal.

\begin{figure}
\includegraphics{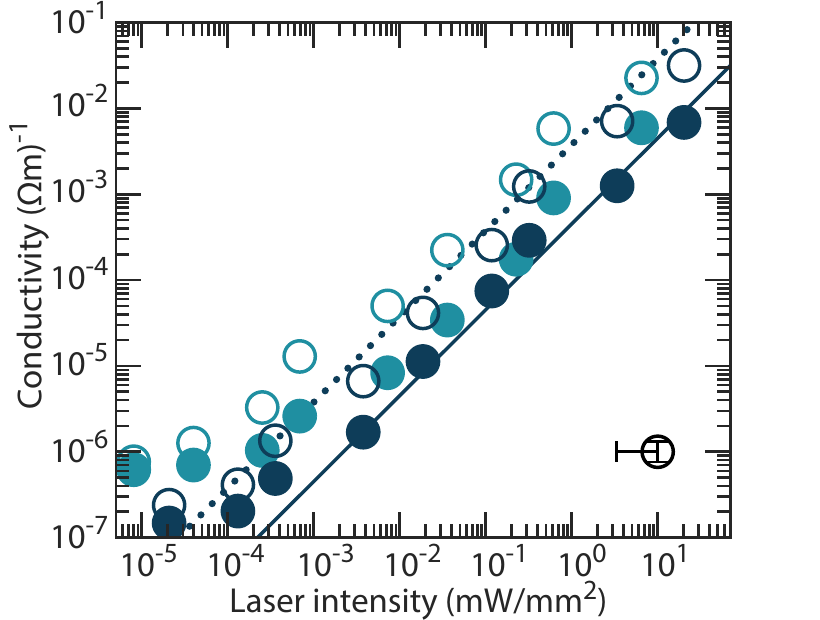}%
\caption{\label{fig:sigmaVpower} The conductivity of the \textsuperscript{28}Si:P with \SI{2e14}{cm^{-3}} (light blue) and \SI{3e15}{cm^{-3}} (dark blue) donors as a function of incident laser power, both on (open dots) and off (filled dots) resonance. The depicted typical error bar (bottom right) indicates the systematic error in delivered laser power due to the unknown absorption of cryostat and resonator windows and the statistical error of the conductivity fit.}
\end{figure}

In Figure \ref{fig:sigmaVpower}, we plot the extracted sample conductivity $\sigma_\mathrm{Si}$ against laser power for two samples with different P concentrations, both on-resonance with the \DX{} transition (open dots) and off-resonance (filled dots). We find a linear relationship between conductivity and laser power over five orders of magnitude, for both samples. The saturation of conductivity for laser powers smaller than \SI{0.1}{\upmu W/mm^2} is likely due to background radiation leaking through the cryostat window. 
The conductivity is consistently larger by 5--8 times under on-resonant illumination compared to off-resonant illumination.

We first discuss the origin of conductivity when illuminating on-resonance with the \DX{} transition. 
Using the known oscillator strength \cite{Dean1967} and the measured inhomogenous linewidth (\SI{1}{GHz} at \SI{0}{mT}) of the \DX{} transition we can estimate the steady-state \DX{} carrier generation rate $G_\mathrm{D^0X}$, which is linearly dependent on the laser intensity $I_L$ (see Appendix). The steady state photocarrier density is then given by $n=G_\mathrm{D^0X}\tau_n$, where $\tau_n$ is the carrier lifetime, such that the conductivity under illumination is $\sigma = e \mu_n n = eG_\mathrm{D^0X} \mu_n \tau_n$. 
From the linear dependence of conductivity with laser power we thus deduce that both $\tau_n$ and $\mu_n$ are approximately constant for the laser intensities studied here.
Furthermore, taking a silicon mobility of $\mu_n  \sim \SI{ 7e4}{cm^2(Vs)^{-1}}$ appropriate at this temperature for these donor and acceptor concentrations~\cite{Norton1973}, the slope in Figure \ref{fig:sigmaVpower} (dotted line) can be fit to give $\tau_n \approx\SI{7}{ns}$. 
The photo-carrier lifetime is expected to be limited by a capture process of the conduction band electron by an ionised donor. The constant value of the carrier lifetime which we observe can be understood by considering the significant boron concentration in this material (in the region of $10^{14}$~cm$^{-3}$), which results in a substantial ionised donor concentration, even in the absence of any illumination. Indeed, the carrier concentration $n$ is estimated to be $5\times10^{10}$~cm$^{-3}$ for $I_\mathrm{L}=10$~mW/mm$^2$ (see Appendix), such that the optically-induced ionised donor concentration is negligible compared to that arising from the compensation in the material.
Using the capture recombination coefficient \cite{Sclar1984a,Sclar1984b} $B_{N^+}=\SI{6.9e-6}{cm^3s^{-1}}$, we infer a constant ionized donor concentration of $N^+ = \SI{\sim 2e13}{cm^{-3}}$ during illumination (see Appendix), consistent with the sample boron concentration~\cite{Dirksen1989}. 

The origin of the enhanced sample conductivity under off-resonant laser illumination
is most likely the direct ionisation of donors into the continuum of the conduction band, creating high energy (hot) electrons. The ionization cross-section for this process \cite{Sclar1984b} is on the order of $\sigma_\mathrm{N^0\rightarrow N^+} \approx 10^{-16}$~cm$^{2}$, which would result in the solid blue line shown in Fig.\ \ref{fig:sigmaVpower}
(following similar arguments to those given above and using the same $\tau_n \approx \SI{7}{ns}$). In this way, both the on-resonant and off-resonant signal can be explained using known values for the generation rate and a common photocarrier lifetime. 
For completeness, a second mechanism that could produce similar observed behaviour is phonon-assisted excitation across the band-gap. The photon-energy at \SI{1078}{nm} is below the silicon band-gap~\cite{Cardona2004}, requiring the absorption of a phonon for such across-gap excitation. However, at low temperatures the phonon bath is frozen out, leading to very small absorption coefficients below the band-gap \cite{Macfarlane1958, Rajkanan1979}.

The ratio of on- to off-resonant conductivity is between 5--8 for both donor concentrations studied here.  This ratio is given by the relative magnitude of the \DX{} generation rate versus the direct ionisation rate of a donor.
%
%
An outstanding question is the fact that the photoconductivity and laser intensity dependence we measure for two samples [P] = \SI{2e14}{cm^{-3}} and \SI{3e15}{cm^{-3}} are similar (see Fig.\ \ref{fig:sigmaVpower}). The lower carrier generation rate expected for the sample with lower donor concentration may be somewhat compensated by a larger mobility and longer carrier lifetime, and may also be influenced by relatively small differences in the boron concentration (and thus ionised donor concentration) for which precise values are not known in these samples.
A further study with a wider range of samples would be required to explore this in more detail.

\begin{figure*}[t]
\includegraphics[width=\textwidth]{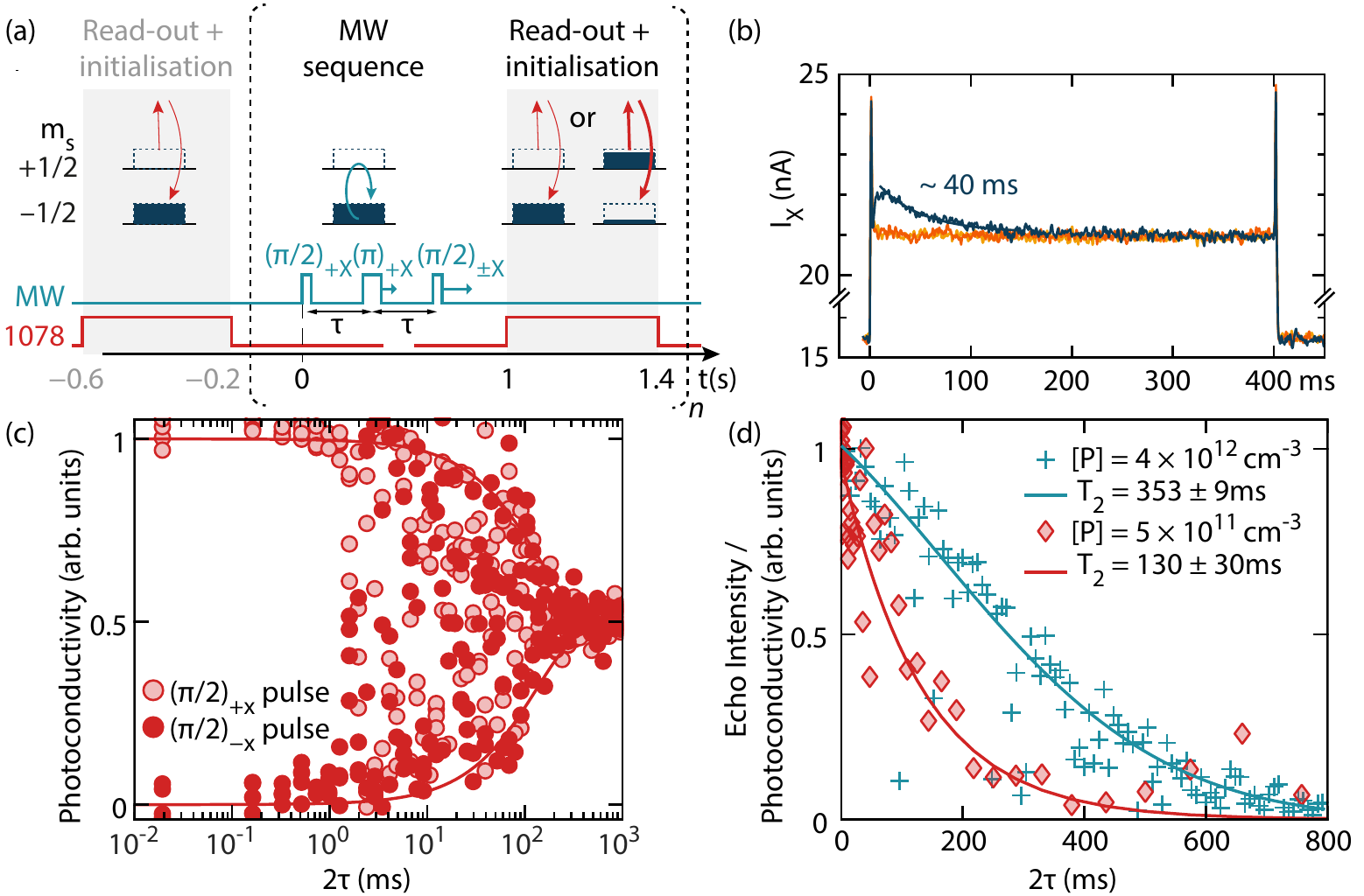}%
\caption{\label{fig:T2} The pulse sequence and results of the photoconductive coherence time measurement. (a) Applied control sequence, consisting of an initialisation laser pulse (also serving as the read-out pulse for the previous experiment), the microwave (mw) pulse sequence and the read-out laser pulse (also serving as initialisation for the next experiment). (b) The photocurrent time trace during the readout laser pulse when applying a single microwave $\pi$ pulse with (i) $B$ on-resonance (dark blue), (ii) $B$ off-resonance (orange) and (iii) $B$ on-resonance with no microwave pulse (yellow). (c) 
Two-pulse echo decay measured by integrating a photocurrent during the read-out laser pulse as in (b) plotted against the interpulse delay, $2\tau$ (sample [P]$=\SI{5e11}{cm^{-3}}$ at 4.5~K.
(d) 
The selected (maximum) data points from (c) and accompanying exponential fit to the data indicating \ttwo $= 130$~ms.  Also shown is the Hahn echo decay measured for [P]$=\SI{4e12}{cm^{-3}}$ at 1.7~K using a conventional pulsed ESR with only one laser pulse for spin initialisation (and not for read-out).}
\end{figure*}

\section{Coherence time}
Having characterised the change in sample conductivity upon excitation of the  \DX{} transition, we now use this as a readout mechanism to measure electron spin coherence times in \sitwoeight\ material with donor concentration below the sensitivity limits of conventional ESR.
([P]~$= \SI{5e11}{cm^{-3}}$, [B]~$=10^{13}$~cm$^{-3}$).
We used a sequence of microwave (X-band, 9.75~GHz) and laser pulses, shown in Fig.~\ref{fig:T2}(a), with the laser tuned to the $m_s=+1/2 : m_h=+1/2$ \DX{} transition (see Fig.\ \ref{fig:Setup}), addressing the spin \spinup\ ground state, and a magnetic field of around 347~mT. 
Through optical pumping, the first laser pulse of \SI{400}{ms} therefore polarizes and initializes the donor electron into the spin \spindown\ state. A microwave Hahn echo pulse sequence follows, with a $(+/-)\pi/2$ pulse applied at the time of the electron spin echo formation to project the refocused coherent electron spin state into \spinup\ or \spindown, respectively. The final `read out' laser pulse creates a transient conductivity signal which depends on  the \spinup\ population remaining at the end of the microwave pulse sequence. The time-resolved sample conductivity is measured using a lock-in amplifier (Stanford Research Systems SR830, $V_\mathrm{AC}=\SI{1}{V}$, $f\approx20$~kHz), whose phase sensitive current output is captured on an oscilloscope. As seen in Fig.~\ref{fig:T2}(b)), only if some donor \spinup\ population has been generated by the microwave pulse sequence is a distinct conductivity transient observed during the `read-out' pulse --- this originates from the Auger electrons produced following laser-induced \DX\ generation. The transient decays with a time constant of \SI{\sim 40}{ms}, characteristic of the \DX\ excitation rate for our \DX\ linewidth (\SI{200}{MHz}) 
and laser intensity (\SI{0.2}{mW/mm^2}). We maximized the signal by adjusting the lock-in phase and integrated over the first \SI{30}{ms} of transient photoconductivity response to produce a unitless measure for the population of the \spinup\ state, normalising the result using that measured in the same experiment with a short (10~$\mu$s) evolution time.

Figure \ref{fig:T2}(c) plots this measured \spinup\ population as a function of free evolution time, $2\tau$, in the Hahn echo experiment, providing a measure of the electron spin coherence time. 
Three distinct time scales can be observed in the evolution:
For $2\tau\lesssim 1$ ms, the microwave $\pm\pi/2$ pulses consistently project the echo signal into the opposite spin states (\spinup\ and \spindown), as expected, indicating negligible decay in electron spin coherence on this timescale. 
For $\SI{1}{ms} \lesssim 2\tau \lesssim \SI{100}{ms}$ the echo signal appears randomly projected between the \spinup\ and \spindown states, indicating there is a macroscopic coherent state across the electron spin ensemble but its phase varies randomly from one measurement to the next.
This `phase-noise' effect is commonly observed in ESR electromagnets for $ 2\tau \gtrsim \SI{1}{ms}$ and can be attributed to fluctuations of the external magnetic field on a time scale of \SI{\sim 1}{kHz} which impact the net phase acquired by the spin ensemble during an experiment~\cite{TyryshkinJPCM2006}. In a conventional ESR measurement employing detection of both quadrature channels, the effects of such phase noise can be mitigated by recording the magnitude across the two quadrature channels for each experimental shot~\cite{TyryshkinJPCM2006}. In contrast, when using a projective read-out method (such as the photoconductivity measurement used here), the effects of phase noise are manifest as random projections into the \spinup{} and \spindown{} states, as we observe. 
Nevertheless, the maximum value of these randomly projected states can, with a sufficient number of measurements, provide a reasonable measure of the overall echo intensity~\cite{Saeedi2013}. This phase noise is believed to be limited by the instrumentation  and could in principle be improved with superconducting magnets in persistent mode or permanent magnets with magnetic shielding~\cite{Ruster2016}.
For $2\tau > \SI{100}{ms}$ the echo intensity collapses due to electron spin decoherence. The timescale of this collapse can be extracted from a fit to the maximum data points (red diamonds in Fig.\ \ref{fig:T2}(d)) and we find a coherence time (\ttwo) of $130 \pm 30~\si{\milli\second}$. 

The \ttwo\ we measure above is somewhat shorter than expected for this sample. Instantaneous diffusion is not expected to play a role since the concentration is too low~\cite{Tyryshkin2012,Kurshev1992} ($T_\mathrm{2,id} \approx \SI{5}{s}$ for $\mathrm{[P] = \SI{5e11}{cm^{-3}}}$). The coherence time is not limited by a $T_1$ process or by donor-acceptor recombination \cite{Dirksen1989}, as an inversion recovery measurement gives a lower bound for such processes as $T_1>\SI{2}{s}$. Instead, we expect the intrinsic $T_2$ to be limited by spectral diffusion from the residual 47\,ppm of \textsuperscript{29}Si nuclear spins at $T_\mathrm{2,sd} \approx$ 0.5--1~s \cite{Abe2010, Witzel2010}.  The apparent discrepancy between this and the measured value is likely due to instrument limitations, in particular sample vibrations within an inhomogeneous magnetic field (caused for example by the gas flow in the He cryostat). 
Indeed, a suggestive `revival' in spin coherence can be observed in the data of Fig.\ \ref{fig:T2}(d)) around 650~ms. While such features are not fully understood, we have observed similar revivals in other measurement when limited by magnetic field noise and believe them to be a consequence of the specific inhomogeneous magnetic field noise spectrum (which varies according to cryostat, sample mount, magnet power supply, etc.). If this interpretation is correct, the coherence time in the sample inferred from the magnitude of the revival would lie in the expected range of 0.5--1~s.

%
%
To explore this hypothesis further, we compare the results above with measurements of coherence times observed on a higher doped sample ([P]~$=\SI{4e12}{cm^{-3}}$ submerged in superfluid helium at \SI{1.7}{K}, and detected by conventional ESR combined with optical hyperpolarisation via the \DX\ transition (using a 200~ms laser pulse). Here, vibrations are significantly reduced and a coherence time of \SI{350}{ms} was measured (see Fig.\ \ref{fig:T2}(d)), consistent with the product of a known instantaneous diffusion term ($T_\mathrm{2,id}= \SI{600}{ms}$~\cite{Kurshev1992}), and a fitted spectral diffusion term $T_\mathrm{2,sd} = \SI{530}{ms}$, in accordance with literature values for the nuclear spin spectral diffusion of 47\,ppm \textsuperscript{29}Si    \cite{Abe2010}, 
\footnote{Another difference between the two \ttwo~measurements is the delay time between the end of the polarizing laser pulse and beginning of the Hahn echo sequence (\SI{900}{ms} for the echo-detected experiments and 200--450~ms for the photoconductivity-detected ESR experiments). Charge reconfiguration and resulting Stark-field noise during the dark period could induce decoherence, however, while these effects have not been studied systematically, no difference in photoconductivity-measured $T_2$ has been observed for the two delay times \SI{200}{ms} and \SI{450}{ms}.}.

\section{Conclusions}

In summary, we used contactless capacitive measurements to characterize the photoconductivity of doped silicon samples under resonant \DX{} excitation. We have shown how the spin-dependent \DX{} photoconductivity can be used as a method to detect pulsed ESR on samples with doping levels below the sensitivity of conventional ESR, with a single shot signal-to-noise ratio of about 10 for spin ensembles with concentration \SI{5e11}{cm^{-3}}. However, efforts must be taken to minimise sample vibrations in order to observe coherence times on the timescale of seconds or longer. 
We find evidence that off-resonant contributions to photo-conductivity arise primarily from direct ionization of donors. As a result this photoconductivity-detected ESR scales down to much  smaller ensembles because the ratio of on- to off-resonant photo conductivity, a key factor in the signal-to-noise ratio, would be independent of the donor ensemble size.



\section{Acknowledgements}
The $^{28}$Si samples used in this study were prepared from Avo28 crystal produced by the International Avogadro Coordination (IAC) Project (2004-2011) in cooperation among the BIPM, the INRIM (Italy), the IRMM (EU), the NMIA (Australia), the NMIJ (Japan), the NPL (UK), and the PTB (Germany). This research was supported by the Engineering and Physical Sciences Research Council (EPSRC) through UNDEDD (EP/K025945/1) and a Doctoral Training Grant, as well as by the European UnionÃs Horizon 2020 research and innovatioN programme under Grant Agreement Nos. 688539 (MOS-QUITO) and 771493 (LOQO-MOTIONS). The work at Princeton was supported by the NSF MRSEC Program (Grant No. DMR-1420541) and the ARO (Grant No. W911NF-13-1-0179).

\bibliography{arxivv3}

\newpage
\newpage
\pagebreak
\clearpage
\appendix

%
%

\section{Electron carrier generation rates under illumination}
\label{ch:App1}

The on-resonance carrier generation rate $G_\mathrm{D^0X}$ can be calculated from the Einstein coefficient $B_{12}^f$ (related to the oscillator strength \cite{Schmid1977}) of the \DX{} transition and its measured lineshape function $g(f,f_0)$ as: 
\begin{equation}
G_\mathrm{D^0X} =  N^0 B_{12}^f \int_0^\infty g(f,f_0) \rho(f) df
\end{equation}
Here $N^0$ is the density of neutral donors and  $\rho(f) = I_\mathrm{L}/(c/n_\mathrm{Si}) \delta(f_0)$ is the $\delta$-like power spectral density  of the laser within the silicon sample due to a laser with intensity $I_\mathrm{L}$. The linewidth of the laser is much smaller than the  \DX{} linewidth $\Delta f$, and hence the integral yields:

\begin{equation}
G_\mathrm{D^0X} =  N^0 B_{12}^f  \frac{I_\mathrm{L}}{c/n_\mathrm{Si}} \frac{2}{\pi \Delta f}
\end{equation}

The carrier generation rate due to resonant bound exciton generation is thus proportional to the laser intensity, the donor density and inversely proportional to the \DX{} linewidth. $ B_{12}= 3.1 \times 10^{16}$~J$^{-1}$m$^3$s$^{-2}$ for P and $c/n_\mathrm{Si}$ is the speed of light in silicon $= 8.1\times 10^7$ms$^{-1}$.

Off resonance, the direct ionisation rate of the laser --- assumed to be the main cause of  extrinsic photoconductivity --- scales with the capture cross-section $\sigma_\mathrm{N^0\rightarrow N^+} $ according to  \cite{Bube1992}:
\begin{equation}
G_\mathrm{direct} = N^0 \frac{I_\mathrm{L}}{\hbar\omega} \sigma_\mathrm{N^0\rightarrow N^+} 
\end{equation}
Data for $\sigma_\mathrm{N^0\rightarrow N^+} $  of phosphorus at a wavelength of \SI{1078}{nm} are scarce, but can be extrapolated from the figures presented in \cite{Sclar1984b}. By comparing the two equations, it can be observed that the ratio of on- to off-resonant excitation is  independent of donor density and laser intensity.

Finally, we present a few remarks regarding the band-to-band excitation of electrons from the valence band into the conduction band. As discussed in the main text, the required phonon-absorption is heavily surpressed due to the low temperature and hence very small absorption coefficients are measured for silicon below the band-gap and at low temperatures \cite{Macfarlane1958}.  Data close to the band-gap are scarce but unpublished measurements of R.Nawrodt (Friedrich Schiller University Jena) give an upper limit for the intrinsic absorption of $\alpha_\mathrm{Si}(\SI{3}{K}, \SI{1.15}{eV}) \approx \SI{3e-4}{\per\centi\meter}$, although the exact origin of the absorption for this wavelength remains unclear and could still be of extrinsic origin. Still, the  carrier generation rate associated with this upper limit for intrinsic absorption is much smaller than the expected $G_\mathrm{direct} $ at the donor densities studied here. We thus do not expect intrinsic absorption to contribute to the photoconductivities presented in this Letter.


\end{document}

%% file: defs_thesis.tex
\newcommand {\spinup} {$\left|\uparrow\right\rangle$}
\newcommand {\spindown} {$\left|\downarrow\right\rangle$}

\newcommand{\beq}{\begin{equation}}
\newcommand{\eeq}{\end{equation}}

\newcommand{\sitwoeight}{$^{28}$Si}

\newcommand*{\DX}{D\textsuperscript{0}X}

\newcommand{\beqa}{\begin{eqnarray}}
\newcommand{\eeqa}{\end{eqnarray}}

\newcommand{\ttwo}{$T_2$}

%% file: arxivv3.bbl
\begin{thebibliography}{33}%
\makeatletter
\providecommand \@ifxundefined [1]{%
 \@ifx{#1\undefined}
}%
\providecommand \@ifnum [1]{%
 \ifnum #1\expandafter \@firstoftwo
 \else \expandafter \@secondoftwo
 \fi
}%
\providecommand \@ifx [1]{%
 \ifx #1\expandafter \@firstoftwo
 \else \expandafter \@secondoftwo
 \fi
}%
\providecommand \natexlab [1]{#1}%
\providecommand \enquote  [1]{``#1''}%
\providecommand \bibnamefont  [1]{#1}%
\providecommand \bibfnamefont [1]{#1}%
\providecommand \citenamefont [1]{#1}%
\providecommand \href@noop [0]{\@secondoftwo}%
\providecommand \href [0]{\begingroup \@sanitize@url \@href}%
\providecommand \@href[1]{\@@startlink{#1}\@@href}%
\providecommand \@@href[1]{\endgroup#1\@@endlink}%
\providecommand \@sanitize@url [0]{\catcode `\\12\catcode `\$12\catcode
  `\&12\catcode `\#12\catcode `\^12\catcode `\_12\catcode `\%12\relax}%
\providecommand \@@startlink[1]{}%
\providecommand \@@endlink[0]{}%
\providecommand \url  [0]{\begingroup\@sanitize@url \@url }%
\providecommand \@url [1]{\endgroup\@href {#1}{\urlprefix }}%
\providecommand \urlprefix  [0]{URL }%
\providecommand \Eprint [0]{\href }%
\providecommand \doibase [0]{http://dx.doi.org/}%
\providecommand \selectlanguage [0]{\@gobble}%
\providecommand \bibinfo  [0]{\@secondoftwo}%
\providecommand \bibfield  [0]{\@secondoftwo}%
\providecommand \translation [1]{[#1]}%
\providecommand \BibitemOpen [0]{}%
\providecommand \bibitemStop [0]{}%
\providecommand \bibitemNoStop [0]{.\EOS\space}%
\providecommand \EOS [0]{\spacefactor3000\relax}%
\providecommand \BibitemShut  [1]{\csname bibitem#1\endcsname}%
\let\auto@bib@innerbib\@empty
\bibitem [{\citenamefont {Becker}\ \emph {et~al.}(2010)\citenamefont {Becker},
  \citenamefont {Pohl}, \citenamefont {Riemann},\ and\ \citenamefont
  {Abrosimov}}]{Becker2010}%
  \BibitemOpen
  \bibfield  {author} {\bibinfo {author} {\bibfnamefont {P.}~\bibnamefont
  {Becker}}, \bibinfo {author} {\bibfnamefont {H.~J.}\ \bibnamefont {Pohl}},
  \bibinfo {author} {\bibfnamefont {H.}~\bibnamefont {Riemann}}, \ and\
  \bibinfo {author} {\bibfnamefont {N.}~\bibnamefont {Abrosimov}},\ }\bibfield
  {title} {\enquote {\bibinfo {title} {{Enrichment of silicon for a better
  kilogram}},}\ }\href {\doibase 10.1002/pssa.200925148} {\bibfield  {journal}
  {\bibinfo  {journal} {Physica Status Solidi (a)}\ }\textbf {\bibinfo {volume}
  {207}},\ \bibinfo {pages} {49--66} (\bibinfo {year} {2010})}\BibitemShut
  {NoStop}%
\bibitem [{\citenamefont {Steger}\ \emph {et~al.}(2012)\citenamefont {Steger},
  \citenamefont {Saeedi}, \citenamefont {Thewalt}, \citenamefont {Morton},
  \citenamefont {Riemann}, \citenamefont {Abrosimov}, \citenamefont {Becker},\
  and\ \citenamefont {Pohl}}]{Steger2012}%
  \BibitemOpen
  \bibfield  {author} {\bibinfo {author} {\bibfnamefont {M.}~\bibnamefont
  {Steger}}, \bibinfo {author} {\bibfnamefont {K.}~\bibnamefont {Saeedi}},
  \bibinfo {author} {\bibfnamefont {M.~L.~W.}\ \bibnamefont {Thewalt}},
  \bibinfo {author} {\bibfnamefont {J.~J.~L.}\ \bibnamefont {Morton}}, \bibinfo
  {author} {\bibfnamefont {H.}~\bibnamefont {Riemann}}, \bibinfo {author}
  {\bibfnamefont {N.~V.}\ \bibnamefont {Abrosimov}}, \bibinfo {author}
  {\bibfnamefont {P.}~\bibnamefont {Becker}}, \ and\ \bibinfo {author}
  {\bibfnamefont {H.-J.}\ \bibnamefont {Pohl}},\ }\bibfield  {title} {\enquote
  {\bibinfo {title} {{Quantum information storage for over 180 s using donor
  spins in a 28Si ``semiconductor vacuum''}},}\ }\href
  {http://www.sciencemag.org/content/336/6086/1280.short} {\bibfield  {journal}
  {\bibinfo  {journal} {Science}\ }\textbf {\bibinfo {volume} {336}},\ \bibinfo
  {pages} {6086} (\bibinfo {year} {2012})}\BibitemShut {NoStop}%
\bibitem [{\citenamefont {Saeedi}\ \emph {et~al.}(2013)\citenamefont {Saeedi},
  \citenamefont {Simmons}, \citenamefont {Salvail}, \citenamefont {Dluhy},
  \citenamefont {Riemann}, \citenamefont {Abrosimov}, \citenamefont {Becker},
  \citenamefont {Pohl}, \citenamefont {Morton},\ and\ \citenamefont
  {Thewalt}}]{Saeedi2013}%
  \BibitemOpen
  \bibfield  {author} {\bibinfo {author} {\bibfnamefont {K.}~\bibnamefont
  {Saeedi}}, \bibinfo {author} {\bibfnamefont {S.}~\bibnamefont {Simmons}},
  \bibinfo {author} {\bibfnamefont {J.~Z.}\ \bibnamefont {Salvail}}, \bibinfo
  {author} {\bibfnamefont {P.}~\bibnamefont {Dluhy}}, \bibinfo {author}
  {\bibfnamefont {H.}~\bibnamefont {Riemann}}, \bibinfo {author} {\bibfnamefont
  {N.~V.}\ \bibnamefont {Abrosimov}}, \bibinfo {author} {\bibfnamefont
  {P.}~\bibnamefont {Becker}}, \bibinfo {author} {\bibfnamefont {H.-J.}\
  \bibnamefont {Pohl}}, \bibinfo {author} {\bibfnamefont {J.~J.~L.}\
  \bibnamefont {Morton}}, \ and\ \bibinfo {author} {\bibfnamefont {M.~L.~W.}\
  \bibnamefont {Thewalt}},\ }\bibfield  {title} {\enquote {\bibinfo {title}
  {Room-temperature quantum bit storage exceeding 39 minutes using ionized
  donors in silicon-28},}\ }\href {\doibase 10.1126/science.1239584} {\bibfield
   {journal} {\bibinfo  {journal} {Science}\ }\textbf {\bibinfo {volume}
  {342}},\ \bibinfo {pages} {830--833} (\bibinfo {year} {2013})}\BibitemShut
  {NoStop}%
\bibitem [{\citenamefont {Tyryshkin}\ \emph {et~al.}(2012)\citenamefont
  {Tyryshkin}, \citenamefont {Tojo}, \citenamefont {Morton}, \citenamefont
  {Riemann}, \citenamefont {Abrosimov}, \citenamefont {Becker}, \citenamefont
  {Pohl}, \citenamefont {Schenkel}, \citenamefont {Thewalt}, \citenamefont
  {Itoh},\ and\ \citenamefont {Lyon}}]{Tyryshkin2012}%
  \BibitemOpen
  \bibfield  {author} {\bibinfo {author} {\bibfnamefont {A.~M.}\ \bibnamefont
  {Tyryshkin}}, \bibinfo {author} {\bibfnamefont {S.}~\bibnamefont {Tojo}},
  \bibinfo {author} {\bibfnamefont {J.~J.~L.}\ \bibnamefont {Morton}}, \bibinfo
  {author} {\bibfnamefont {H.}~\bibnamefont {Riemann}}, \bibinfo {author}
  {\bibfnamefont {N.~V.}\ \bibnamefont {Abrosimov}}, \bibinfo {author}
  {\bibfnamefont {P.}~\bibnamefont {Becker}}, \bibinfo {author} {\bibfnamefont
  {H.-J.}\ \bibnamefont {Pohl}}, \bibinfo {author} {\bibfnamefont
  {T.}~\bibnamefont {Schenkel}}, \bibinfo {author} {\bibfnamefont {M.~L.~W.}\
  \bibnamefont {Thewalt}}, \bibinfo {author} {\bibfnamefont {K.~M.}\
  \bibnamefont {Itoh}}, \ and\ \bibinfo {author} {\bibfnamefont {S.~A.}\
  \bibnamefont {Lyon}},\ }\bibfield  {title} {\enquote {\bibinfo {title}
  {{Electron spin coherence exceeding seconds in high-purity silicon}},}\
  }\href {http://dx.doi.org/10.1038/nmat3182} {\bibfield  {journal} {\bibinfo
  {journal} {Nature Materials}\ }\textbf {\bibinfo {volume} {11}},\ \bibinfo
  {pages} {143--147} (\bibinfo {year} {2012})}\BibitemShut {NoStop}%
\bibitem [{\citenamefont {Wolfowicz}\ \emph {et~al.}(2013)\citenamefont
  {Wolfowicz}, \citenamefont {Tyryshkin}, \citenamefont {George}, \citenamefont
  {Riemann}, \citenamefont {Abrosimov}, \citenamefont {Becker}, \citenamefont
  {Pohl}, \citenamefont {Thewalt}, \citenamefont {Lyon},\ and\ \citenamefont
  {Morton}}]{Wolfowicz2013}%
  \BibitemOpen
  \bibfield  {author} {\bibinfo {author} {\bibfnamefont {Gary}\ \bibnamefont
  {Wolfowicz}}, \bibinfo {author} {\bibfnamefont {Alexei~M}\ \bibnamefont
  {Tyryshkin}}, \bibinfo {author} {\bibfnamefont {Richard~E}\ \bibnamefont
  {George}}, \bibinfo {author} {\bibfnamefont {Helge}\ \bibnamefont {Riemann}},
  \bibinfo {author} {\bibfnamefont {Nikolai~V}\ \bibnamefont {Abrosimov}},
  \bibinfo {author} {\bibfnamefont {Peter}\ \bibnamefont {Becker}}, \bibinfo
  {author} {\bibfnamefont {Hans-Joachim}\ \bibnamefont {Pohl}}, \bibinfo
  {author} {\bibfnamefont {Mike L~W}\ \bibnamefont {Thewalt}}, \bibinfo
  {author} {\bibfnamefont {Stephen~a}\ \bibnamefont {Lyon}}, \ and\ \bibinfo
  {author} {\bibfnamefont {John J~L}\ \bibnamefont {Morton}},\ }\bibfield
  {title} {\enquote {\bibinfo {title} {{Atomic clock transitions in
  silicon-based spin qubits.}}}\ }\href {\doibase 10.1038/nnano.2013.117}
  {\bibfield  {journal} {\bibinfo  {journal} {Nature Nanotechnology}\ }\textbf
  {\bibinfo {volume} {8}},\ \bibinfo {pages} {561--4} (\bibinfo {year}
  {2013})}\BibitemShut {NoStop}%
\bibitem [{\citenamefont {Muhonen}\ \emph {et~al.}(2014)\citenamefont
  {Muhonen}, \citenamefont {Dehollain}, \citenamefont {Laucht}, \citenamefont
  {Hudson}, \citenamefont {Kalra}, \citenamefont {Sekiguchi}, \citenamefont
  {Itoh}, \citenamefont {Jamieson}, \citenamefont {McCallum}, \citenamefont
  {Dzurak},\ and\ \citenamefont {Morello}}]{Muhonen2014}%
  \BibitemOpen
  \bibfield  {author} {\bibinfo {author} {\bibfnamefont {Juha~T.}\ \bibnamefont
  {Muhonen}}, \bibinfo {author} {\bibfnamefont {Juan~P.}\ \bibnamefont
  {Dehollain}}, \bibinfo {author} {\bibfnamefont {Arne}\ \bibnamefont
  {Laucht}}, \bibinfo {author} {\bibfnamefont {Fay~E.}\ \bibnamefont {Hudson}},
  \bibinfo {author} {\bibfnamefont {Rachpon}\ \bibnamefont {Kalra}}, \bibinfo
  {author} {\bibfnamefont {Takeharu}\ \bibnamefont {Sekiguchi}}, \bibinfo
  {author} {\bibfnamefont {Kohei~M.}\ \bibnamefont {Itoh}}, \bibinfo {author}
  {\bibfnamefont {David~N.}\ \bibnamefont {Jamieson}}, \bibinfo {author}
  {\bibfnamefont {Jeffrey~C.}\ \bibnamefont {McCallum}}, \bibinfo {author}
  {\bibfnamefont {Andrew~S.}\ \bibnamefont {Dzurak}}, \ and\ \bibinfo {author}
  {\bibfnamefont {Andrea}\ \bibnamefont {Morello}},\ }\bibfield  {title}
  {\enquote {\bibinfo {title} {Storing quantum information for 30 seconds in a
  nanoelectronic device},}\ }\href {http://dx.doi.org/10.1038/nnano.2014.211}
  {\bibfield  {journal} {\bibinfo  {journal} {Nature Nanotechnology}\ }\textbf
  {\bibinfo {volume} {9}},\ \bibinfo {pages} {986--991} (\bibinfo {year}
  {2014})}\BibitemShut {NoStop}%
\bibitem [{\citenamefont {Kurshev}\ and\ \citenamefont
  {Ichikawa}(1992)}]{Kurshev1992}%
  \BibitemOpen
  \bibfield  {author} {\bibinfo {author} {\bibfnamefont {Vadim~V}\ \bibnamefont
  {Kurshev}}\ and\ \bibinfo {author} {\bibfnamefont {Tsuneki}\ \bibnamefont
  {Ichikawa}},\ }\bibfield  {title} {\enquote {\bibinfo {title} {Effect of spin
  flip-flop on electron-spin-echo decay due to instantaneous diffusion},}\
  }\href {\doibase 10.1016/0022-2364(92)90341-4} {\bibfield  {journal}
  {\bibinfo  {journal} {Journal of Magnetic Resonance (1969)}\ }\textbf
  {\bibinfo {volume} {96}},\ \bibinfo {pages} {563 -- 573} (\bibinfo {year}
  {1992})}\BibitemShut {NoStop}%
\bibitem [{\citenamefont {Schweiger}\ and\ \citenamefont
  {Jeschke}(2001)}]{Schweiger2001}%
  \BibitemOpen
  \bibfield  {author} {\bibinfo {author} {\bibfnamefont {Arthur}\ \bibnamefont
  {Schweiger}}\ and\ \bibinfo {author} {\bibfnamefont {Gunnar}\ \bibnamefont
  {Jeschke}},\ }\href@noop {} {\emph {\bibinfo {title} {Principles of pulse
  electron paramagnetic resonance}}}\ (\bibinfo  {publisher} {Oxford University
  Press},\ \bibinfo {year} {2001})\BibitemShut {NoStop}%
\bibitem [{\citenamefont {Stegner}\ \emph {et~al.}(2006)\citenamefont
  {Stegner}, \citenamefont {Boehme}, \citenamefont {Huebl}, \citenamefont
  {Stutzmann}, \citenamefont {Lips},\ and\ \citenamefont
  {Brandt}}]{Stegner2006}%
  \BibitemOpen
  \bibfield  {author} {\bibinfo {author} {\bibfnamefont {Andre~R.}\
  \bibnamefont {Stegner}}, \bibinfo {author} {\bibfnamefont {Christoph}\
  \bibnamefont {Boehme}}, \bibinfo {author} {\bibfnamefont {Hans}\ \bibnamefont
  {Huebl}}, \bibinfo {author} {\bibfnamefont {Martin}\ \bibnamefont
  {Stutzmann}}, \bibinfo {author} {\bibfnamefont {Klaus}\ \bibnamefont {Lips}},
  \ and\ \bibinfo {author} {\bibfnamefont {Martin~S.}\ \bibnamefont {Brandt}},\
  }\bibfield  {title} {\enquote {\bibinfo {title} {Electrical detection of
  coherent {$^{31}$P} spin quantum states},}\ }\href@noop {} {\bibfield
  {journal} {\bibinfo  {journal} {Nature Physics}\ }\textbf {\bibinfo {volume}
  {2}},\ \bibinfo {pages} {835--838} (\bibinfo {year} {2006})}\BibitemShut
  {NoStop}%
\bibitem [{\citenamefont {McCamey}\ \emph {et~al.}(2006)\citenamefont
  {McCamey}, \citenamefont {Huebl}, \citenamefont {Brandt}, \citenamefont
  {Hutchison}, \citenamefont {McCallum}, \citenamefont {Clark},\ and\
  \citenamefont {Hamilton}}]{McCamey2006}%
  \BibitemOpen
  \bibfield  {author} {\bibinfo {author} {\bibfnamefont {D.~R.}\ \bibnamefont
  {McCamey}}, \bibinfo {author} {\bibfnamefont {H.}~\bibnamefont {Huebl}},
  \bibinfo {author} {\bibfnamefont {M.~S.}\ \bibnamefont {Brandt}}, \bibinfo
  {author} {\bibfnamefont {W.~D.}\ \bibnamefont {Hutchison}}, \bibinfo {author}
  {\bibfnamefont {J.~C.}\ \bibnamefont {McCallum}}, \bibinfo {author}
  {\bibfnamefont {R.~G.}\ \bibnamefont {Clark}}, \ and\ \bibinfo {author}
  {\bibfnamefont {A.~R.}\ \bibnamefont {Hamilton}},\ }\bibfield  {title}
  {\enquote {\bibinfo {title} {Electrically detected magnetic resonance in
  ion-implanted si:p nanostructures},}\ }\href {\doibase 10.1063/1.2358928}
  {\bibfield  {journal} {\bibinfo  {journal} {Applied Physics Letters}\
  }\textbf {\bibinfo {volume} {89}},\ \bibinfo {pages} {182115} (\bibinfo
  {year} {2006})}\BibitemShut {NoStop}%
\bibitem [{\citenamefont {Paik}\ \emph {et~al.}(2010)\citenamefont {Paik},
  \citenamefont {Lee}, \citenamefont {Baker}, \citenamefont {McCamey},\ and\
  \citenamefont {Boehme}}]{Paik2010}%
  \BibitemOpen
  \bibfield  {author} {\bibinfo {author} {\bibfnamefont {S.-Y.}\ \bibnamefont
  {Paik}}, \bibinfo {author} {\bibfnamefont {S.-Y.}\ \bibnamefont {Lee}},
  \bibinfo {author} {\bibfnamefont {W.~J.}\ \bibnamefont {Baker}}, \bibinfo
  {author} {\bibfnamefont {D.~R.}\ \bibnamefont {McCamey}}, \ and\ \bibinfo
  {author} {\bibfnamefont {C.}~\bibnamefont {Boehme}},\ }\bibfield  {title}
  {\enquote {\bibinfo {title} {${T}_{1}$ and ${T}_{2}$ spin relaxation time
  limitations of phosphorous donor electrons near crystalline silicon to
  silicon dioxide interface defects},}\ }\href {\doibase
  10.1103/PhysRevB.81.075214} {\bibfield  {journal} {\bibinfo  {journal} {Phys.
  Rev. B}\ }\textbf {\bibinfo {volume} {81}},\ \bibinfo {pages} {075214}
  (\bibinfo {year} {2010})}\BibitemShut {NoStop}%
\bibitem [{\citenamefont {Lo}\ \emph {et~al.}(2015)\citenamefont {Lo},
  \citenamefont {Urdampilleta}, \citenamefont {Ross}, \citenamefont
  {Gonzalez-Zalba}, \citenamefont {Mansir}, \citenamefont {Lyon}, \citenamefont
  {Thewalt},\ and\ \citenamefont {Morton}}]{Lo2015}%
  \BibitemOpen
  \bibfield  {author} {\bibinfo {author} {\bibfnamefont {C.~C.}\ \bibnamefont
  {Lo}}, \bibinfo {author} {\bibfnamefont {M.}~\bibnamefont {Urdampilleta}},
  \bibinfo {author} {\bibfnamefont {P.}~\bibnamefont {Ross}}, \bibinfo {author}
  {\bibfnamefont {M.~F.}\ \bibnamefont {Gonzalez-Zalba}}, \bibinfo {author}
  {\bibfnamefont {J.}~\bibnamefont {Mansir}}, \bibinfo {author} {\bibfnamefont
  {S.~A.}\ \bibnamefont {Lyon}}, \bibinfo {author} {\bibfnamefont {M.~L.~W.}\
  \bibnamefont {Thewalt}}, \ and\ \bibinfo {author} {\bibfnamefont {J.~J.~L.}\
  \bibnamefont {Morton}},\ }\bibfield  {title} {\enquote {\bibinfo {title}
  {Hybrid optical--electrical detection of donor electron spins with bound
  excitons in silicon},}\ }\href {http://dx.doi.org/10.1038/nmat4250}
  {\bibfield  {journal} {\bibinfo  {journal} {Nature Materials}\ }\textbf
  {\bibinfo {volume} {14}},\ \bibinfo {pages} {490--494} (\bibinfo {year}
  {2015})}\BibitemShut {NoStop}%
\bibitem [{\citenamefont {Yang}\ \emph {et~al.}(2008)\citenamefont {Yang},
  \citenamefont {Steger}, \citenamefont {Lian}, \citenamefont {Thewalt},
  \citenamefont {Uemura}, \citenamefont {Sagara}, \citenamefont {Itoh},
  \citenamefont {Haller}, \citenamefont {Ager}, \citenamefont {Lyon},
  \citenamefont {Konuma},\ and\ \citenamefont {Cardona}}]{Yang2008}%
  \BibitemOpen
  \bibfield  {author} {\bibinfo {author} {\bibfnamefont {A.}~\bibnamefont
  {Yang}}, \bibinfo {author} {\bibfnamefont {M.}~\bibnamefont {Steger}},
  \bibinfo {author} {\bibfnamefont {H.~J.}\ \bibnamefont {Lian}}, \bibinfo
  {author} {\bibfnamefont {M.~L.~W.}\ \bibnamefont {Thewalt}}, \bibinfo
  {author} {\bibfnamefont {M.}~\bibnamefont {Uemura}}, \bibinfo {author}
  {\bibfnamefont {A.}~\bibnamefont {Sagara}}, \bibinfo {author} {\bibfnamefont
  {K.~M.}\ \bibnamefont {Itoh}}, \bibinfo {author} {\bibfnamefont {E.~E.}\
  \bibnamefont {Haller}}, \bibinfo {author} {\bibfnamefont {J.~W.}\
  \bibnamefont {Ager}}, \bibinfo {author} {\bibfnamefont {S.~A.}\ \bibnamefont
  {Lyon}}, \bibinfo {author} {\bibfnamefont {M.}~\bibnamefont {Konuma}}, \ and\
  \bibinfo {author} {\bibfnamefont {M.}~\bibnamefont {Cardona}},\ }\bibfield
  {title} {\enquote {\bibinfo {title} {High-resolution photoluminescence
  measurement of the isotopic-mass dependence of the lattice parameter of
  silicon},}\ }\href {\doibase 10.1103/PhysRevB.77.113203} {\bibfield
  {journal} {\bibinfo  {journal} {Physical Review B}\ }\textbf {\bibinfo
  {volume} {77}},\ \bibinfo {pages} {113203} (\bibinfo {year}
  {2008})}\BibitemShut {NoStop}%
\bibitem [{\citenamefont {Franke}\ \emph {et~al.}(2016)\citenamefont {Franke},
  \citenamefont {Szech}, \citenamefont {Hrubesch}, \citenamefont {Riemann},
  \citenamefont {Abrosimov}, \citenamefont {Becker}, \citenamefont {Pohl},
  \citenamefont {Itoh}, \citenamefont {Thewalt},\ and\ \citenamefont
  {Brandt}}]{Franke2016}%
  \BibitemOpen
  \bibfield  {author} {\bibinfo {author} {\bibfnamefont {David~P.}\
  \bibnamefont {Franke}}, \bibinfo {author} {\bibfnamefont {Michael}\
  \bibnamefont {Szech}}, \bibinfo {author} {\bibfnamefont {Florian~M.}\
  \bibnamefont {Hrubesch}}, \bibinfo {author} {\bibfnamefont {Helge}\
  \bibnamefont {Riemann}}, \bibinfo {author} {\bibfnamefont {Nikolai~V.}\
  \bibnamefont {Abrosimov}}, \bibinfo {author} {\bibfnamefont {Peter}\
  \bibnamefont {Becker}}, \bibinfo {author} {\bibfnamefont {Hans-Joachim}\
  \bibnamefont {Pohl}}, \bibinfo {author} {\bibfnamefont {Kohei~M.}\
  \bibnamefont {Itoh}}, \bibinfo {author} {\bibfnamefont {Michael L.~W.}\
  \bibnamefont {Thewalt}}, \ and\ \bibinfo {author} {\bibfnamefont {Martin~S.}\
  \bibnamefont {Brandt}},\ }\bibfield  {title} {\enquote {\bibinfo {title}
  {Electron nuclear double resonance with donor-bound excitons in silicon},}\
  }\href {\doibase 10.1103/PhysRevB.94.235201} {\bibfield  {journal} {\bibinfo
  {journal} {Phys. Rev. B}\ }\textbf {\bibinfo {volume} {94}},\ \bibinfo
  {pages} {235201} (\bibinfo {year} {2016})}\BibitemShut {NoStop}%
\bibitem [{\citenamefont {Wolfowicz}\ \emph {et~al.}(2014)\citenamefont
  {Wolfowicz}, \citenamefont {Urdampilleta}, \citenamefont {Thewalt},
  \citenamefont {Riemann}, \citenamefont {Abrosimov}, \citenamefont {Becker},
  \citenamefont {Pohl},\ and\ \citenamefont {Morton}}]{Wolfowicz2014}%
  \BibitemOpen
  \bibfield  {author} {\bibinfo {author} {\bibfnamefont {Gary}\ \bibnamefont
  {Wolfowicz}}, \bibinfo {author} {\bibfnamefont {Matias}\ \bibnamefont
  {Urdampilleta}}, \bibinfo {author} {\bibfnamefont {Mike L.~W.}\ \bibnamefont
  {Thewalt}}, \bibinfo {author} {\bibfnamefont {Helge}\ \bibnamefont
  {Riemann}}, \bibinfo {author} {\bibfnamefont {Nikolai~V.}\ \bibnamefont
  {Abrosimov}}, \bibinfo {author} {\bibfnamefont {Peter}\ \bibnamefont
  {Becker}}, \bibinfo {author} {\bibfnamefont {Hans-Joachim}\ \bibnamefont
  {Pohl}}, \ and\ \bibinfo {author} {\bibfnamefont {John J.~L.}\ \bibnamefont
  {Morton}},\ }\bibfield  {title} {\enquote {\bibinfo {title} {Conditional
  control of donor nuclear spins in silicon using stark shifts},}\ }\href
  {\doibase 10.1103/PhysRevLett.113.157601} {\bibfield  {journal} {\bibinfo
  {journal} {Physical Review Letters}\ }\textbf {\bibinfo {volume} {113}},\
  \bibinfo {pages} {157601} (\bibinfo {year} {2014})}\BibitemShut {NoStop}%
\bibitem [{\citenamefont {Schmid}(1977)}]{Schmid1977}%
  \BibitemOpen
  \bibfield  {author} {\bibinfo {author} {\bibfnamefont {W}~\bibnamefont
  {Schmid}},\ }\bibfield  {title} {\enquote {\bibinfo {title} {{Auger lifetimes
  for excitons bound to neutral donors and acceptors in Si}},}\ }\href
  {\doibase 10.1002/pssb.2220840216} {\bibfield  {journal} {\bibinfo  {journal}
  {physica status solidi (b)}\ }\textbf {\bibinfo {volume} {84}},\ \bibinfo
  {pages} {529--540} (\bibinfo {year} {1977})}\BibitemShut {NoStop}%
\bibitem [{\citenamefont {Krupka}\ \emph {et~al.}(2006)\citenamefont {Krupka},
  \citenamefont {Breeze}, \citenamefont {Centeno}, \citenamefont {Alford},
  \citenamefont {Claussen},\ and\ \citenamefont {Jensen}}]{krupka2006}%
  \BibitemOpen
  \bibfield  {author} {\bibinfo {author} {\bibfnamefont {Jerzy}\ \bibnamefont
  {Krupka}}, \bibinfo {author} {\bibfnamefont {Jonathan}\ \bibnamefont
  {Breeze}}, \bibinfo {author} {\bibfnamefont {Anthony}\ \bibnamefont
  {Centeno}}, \bibinfo {author} {\bibfnamefont {Neil}\ \bibnamefont {Alford}},
  \bibinfo {author} {\bibfnamefont {Thomas}\ \bibnamefont {Claussen}}, \ and\
  \bibinfo {author} {\bibfnamefont {Leif}\ \bibnamefont {Jensen}},\ }\bibfield
  {title} {\enquote {\bibinfo {title} {Measurements of permittivity, dielectric
  loss tangent, and resistivity of float-zone silicon at microwave
  frequencies},}\ }\href@noop {} {\bibfield  {journal} {\bibinfo  {journal}
  {IEEE Transactions on Microwave Theory and Techniques}\ }\textbf {\bibinfo
  {volume} {54}},\ \bibinfo {pages} {3995--4001} (\bibinfo {year}
  {2006})}\BibitemShut {NoStop}%
\bibitem [{Note1()}]{Note1}%
  \BibitemOpen
  \bibinfo {note} {The polarisability and concentration of phosphorus is
  negligible compared to the polarisability of the silicon lattice~\cite
  {Dhar1985}}\BibitemShut {NoStop}%
\bibitem [{\citenamefont {Dean}\ \emph {et~al.}(1967)\citenamefont {Dean},
  \citenamefont {Flood},\ and\ \citenamefont {Kaminsky}}]{Dean1967}%
  \BibitemOpen
  \bibfield  {author} {\bibinfo {author} {\bibfnamefont {P.~J.}\ \bibnamefont
  {Dean}}, \bibinfo {author} {\bibfnamefont {W.~F.}\ \bibnamefont {Flood}}, \
  and\ \bibinfo {author} {\bibfnamefont {G.}~\bibnamefont {Kaminsky}},\
  }\bibfield  {title} {\enquote {\bibinfo {title} {{Absorption due to bound
  excitons in silicon}},}\ }\href {\doibase 10.1103/PhysRev.163.721} {\bibfield
   {journal} {\bibinfo  {journal} {Physical Review}\ }\textbf {\bibinfo
  {volume} {163}},\ \bibinfo {pages} {721--725} (\bibinfo {year}
  {1967})}\BibitemShut {NoStop}%
\bibitem [{\citenamefont {Norton}\ \emph {et~al.}(1973)\citenamefont {Norton},
  \citenamefont {Braggins},\ and\ \citenamefont {Levinstein}}]{Norton1973}%
  \BibitemOpen
  \bibfield  {author} {\bibinfo {author} {\bibfnamefont {P.}~\bibnamefont
  {Norton}}, \bibinfo {author} {\bibfnamefont {T.}~\bibnamefont {Braggins}}, \
  and\ \bibinfo {author} {\bibfnamefont {H.}~\bibnamefont {Levinstein}},\
  }\bibfield  {title} {\enquote {\bibinfo {title} {{Impurity and lattice
  scattering parameters as determined from Hall and mobility analysis in n-type
  silicon}},}\ }\href {\doibase 10.1103/PhysRevB.8.5632} {\bibfield  {journal}
  {\bibinfo  {journal} {Physical Review B}\ }\textbf {\bibinfo {volume} {8}},\
  \bibinfo {pages} {5632--5653} (\bibinfo {year} {1973})}\BibitemShut {NoStop}%
\bibitem [{\citenamefont {Sclar}(1984{\natexlab{a}})}]{Sclar1984a}%
  \BibitemOpen
  \bibfield  {author} {\bibinfo {author} {\bibfnamefont {N.}~\bibnamefont
  {Sclar}},\ }\bibfield  {title} {\enquote {\bibinfo {title} {{Properties of
  doped silicon and germanium infrared detectors}},}\ }\href {\doibase
  10.1016/0079-6727(84)90001-6} {\bibfield  {journal} {\bibinfo  {journal}
  {Progress in Quantum Electronics}\ }\textbf {\bibinfo {volume} {9}},\
  \bibinfo {pages} {149--257} (\bibinfo {year}
  {1984}{\natexlab{a}})}\BibitemShut {NoStop}%
\bibitem [{\citenamefont {Sclar}(1984{\natexlab{b}})}]{Sclar1984b}%
  \BibitemOpen
  \bibfield  {author} {\bibinfo {author} {\bibfnamefont {N.}~\bibnamefont
  {Sclar}},\ }\bibfield  {title} {\enquote {\bibinfo {title} {{Asymmetries in
  photoconductive properties of donor and acceptor impurities in silicon}},}\
  }\href {\doibase 10.1063/1.333341} {\bibfield  {journal} {\bibinfo  {journal}
  {Journal of Applied Physics}\ }\textbf {\bibinfo {volume} {55}},\ \bibinfo
  {pages} {2972--2976} (\bibinfo {year} {1984}{\natexlab{b}})}\BibitemShut
  {NoStop}%
\bibitem [{\citenamefont {Dirksen}\ \emph {et~al.}(1989)\citenamefont
  {Dirksen}, \citenamefont {Henstra},\ and\ \citenamefont
  {Wenckebach}}]{Dirksen1989}%
  \BibitemOpen
  \bibfield  {author} {\bibinfo {author} {\bibfnamefont {P}~\bibnamefont
  {Dirksen}}, \bibinfo {author} {\bibfnamefont {A}~\bibnamefont {Henstra}}, \
  and\ \bibinfo {author} {\bibfnamefont {W~Th}\ \bibnamefont {Wenckebach}},\
  }\bibfield  {title} {\enquote {\bibinfo {title} {An electron spin echo study
  of donor-acceptor recombination},}\ }\href
  {http://iopscience.iop.org/0953-8984/1/39/020} {\bibfield  {journal}
  {\bibinfo  {journal} {Journal of Physics: Condensed Matter}\ }\textbf
  {\bibinfo {volume} {1}},\ \bibinfo {pages} {7085} (\bibinfo {year}
  {1989})}\BibitemShut {NoStop}%
\bibitem [{\citenamefont {Cardona}\ \emph {et~al.}(2004)\citenamefont
  {Cardona}, \citenamefont {Meyer},\ and\ \citenamefont
  {Thewalt}}]{Cardona2004}%
  \BibitemOpen
  \bibfield  {author} {\bibinfo {author} {\bibfnamefont {M.}~\bibnamefont
  {Cardona}}, \bibinfo {author} {\bibfnamefont {T.}~\bibnamefont {Meyer}}, \
  and\ \bibinfo {author} {\bibfnamefont {M.~L.~W.}\ \bibnamefont {Thewalt}},\
  }\bibfield  {title} {\enquote {\bibinfo {title} {Temperature dependence of
  the energy gap of semiconductors in the low-temperature limit},}\ }\href
  {\doibase 10.1103/PhysRevLett.92.196403} {\bibfield  {journal} {\bibinfo
  {journal} {Physical Review Letters}\ }\textbf {\bibinfo {volume} {92}},\
  \bibinfo {pages} {196403} (\bibinfo {year} {2004})}\BibitemShut {NoStop}%
\bibitem [{\citenamefont {Macfarlane}\ \emph {et~al.}(1958)\citenamefont
  {Macfarlane}, \citenamefont {McLean}, \citenamefont {Quarrington},\ and\
  \citenamefont {Roberts}}]{Macfarlane1958}%
  \BibitemOpen
  \bibfield  {author} {\bibinfo {author} {\bibfnamefont {GG}~\bibnamefont
  {Macfarlane}}, \bibinfo {author} {\bibfnamefont {TP}~\bibnamefont {McLean}},
  \bibinfo {author} {\bibfnamefont {JE}~\bibnamefont {Quarrington}}, \ and\
  \bibinfo {author} {\bibfnamefont {V}~\bibnamefont {Roberts}},\ }\bibfield
  {title} {\enquote {\bibinfo {title} {Fine structure in the absorption-edge
  spectrum of si},}\ }\href@noop {} {\bibfield  {journal} {\bibinfo  {journal}
  {Physical Review}\ }\textbf {\bibinfo {volume} {111}},\ \bibinfo {pages}
  {1245} (\bibinfo {year} {1958})}\BibitemShut {NoStop}%
\bibitem [{\citenamefont {Rajkanan}\ \emph {et~al.}(1979)\citenamefont
  {Rajkanan}, \citenamefont {Singh},\ and\ \citenamefont
  {Shewchun}}]{Rajkanan1979}%
  \BibitemOpen
  \bibfield  {author} {\bibinfo {author} {\bibfnamefont {K}~\bibnamefont
  {Rajkanan}}, \bibinfo {author} {\bibfnamefont {R}~\bibnamefont {Singh}}, \
  and\ \bibinfo {author} {\bibfnamefont {J}~\bibnamefont {Shewchun}},\
  }\bibfield  {title} {\enquote {\bibinfo {title} {{Absorption coefficient of
  silicon for solar cell calculations}},}\ }\href {\doibase
  http://dx.doi.org/10.1016/0038-1101(79)90128-X} {\bibfield  {journal}
  {\bibinfo  {journal} {Solid-State Electronics}\ }\textbf {\bibinfo {volume}
  {22}},\ \bibinfo {pages} {793--795} (\bibinfo {year} {1979})}\BibitemShut
  {NoStop}%
\bibitem [{\citenamefont {Tyryshkin}\ \emph {et~al.}(2006)\citenamefont
  {Tyryshkin}, \citenamefont {Morton}, \citenamefont {Benjamin}, \citenamefont
  {Ardavan}, \citenamefont {Briggs}, \citenamefont {Ager},\ and\ \citenamefont
  {Lyon}}]{TyryshkinJPCM2006}%
  \BibitemOpen
  \bibfield  {author} {\bibinfo {author} {\bibfnamefont {A.~M.}\ \bibnamefont
  {Tyryshkin}}, \bibinfo {author} {\bibfnamefont {J.~J.~L.}\ \bibnamefont
  {Morton}}, \bibinfo {author} {\bibfnamefont {S.~C.}\ \bibnamefont
  {Benjamin}}, \bibinfo {author} {\bibfnamefont {A.}~\bibnamefont {Ardavan}},
  \bibinfo {author} {\bibfnamefont {G.~A.~D.}\ \bibnamefont {Briggs}}, \bibinfo
  {author} {\bibfnamefont {J.~W.}\ \bibnamefont {Ager}}, \ and\ \bibinfo
  {author} {\bibfnamefont {S.~A.}\ \bibnamefont {Lyon}},\ }\bibfield  {title}
  {\enquote {\bibinfo {title} {Coherence of spin qubits in silicon},}\
  }\href@noop {} {\bibfield  {journal} {\bibinfo  {journal} {J. Phys. Cond.
  Mat.}\ }\textbf {\bibinfo {volume} {18}},\ \bibinfo {pages} {S783--S794}
  (\bibinfo {year} {2006})}\BibitemShut {NoStop}%
\bibitem [{\citenamefont {{Ruster}}\ \emph {et~al.}(2016)\citenamefont
  {{Ruster}}, \citenamefont {{Schmiegelow}}, \citenamefont {{Kaufmann}},
  \citenamefont {{Warschburger}}, \citenamefont {{Schmidt-Kaler}},\ and\
  \citenamefont {{Poschinger}}}]{Ruster2016}%
  \BibitemOpen
  \bibfield  {author} {\bibinfo {author} {\bibfnamefont {T.}~\bibnamefont
  {{Ruster}}}, \bibinfo {author} {\bibfnamefont {C.~T.}\ \bibnamefont
  {{Schmiegelow}}}, \bibinfo {author} {\bibfnamefont {H.}~\bibnamefont
  {{Kaufmann}}}, \bibinfo {author} {\bibfnamefont {C.}~\bibnamefont
  {{Warschburger}}}, \bibinfo {author} {\bibfnamefont {F.}~\bibnamefont
  {{Schmidt-Kaler}}}, \ and\ \bibinfo {author} {\bibfnamefont {U.~G.}\
  \bibnamefont {{Poschinger}}},\ }\bibfield  {title} {\enquote {\bibinfo
  {title} {{A long-lived Zeeman trapped-ion qubit}},}\ }\href {\doibase
  10.1007/s00340-016-6527-4} {\bibfield  {journal} {\bibinfo  {journal}
  {Applied Physics B: Lasers and Optics}\ }\textbf {\bibinfo {volume} {122}},\
  \bibinfo {eid} {254} (\bibinfo {year} {2016})}\BibitemShut {NoStop}%
\bibitem [{\citenamefont {Abe}\ \emph {et~al.}(2010)\citenamefont {Abe},
  \citenamefont {Tyryshkin}, \citenamefont {Tojo}, \citenamefont {Morton},
  \citenamefont {Witzel}, \citenamefont {Fujimoto}, \citenamefont {Ager},
  \citenamefont {Haller}, \citenamefont {Isoya}, \citenamefont {Lyon},
  \citenamefont {Thewalt},\ and\ \citenamefont {Itoh}}]{Abe2010}%
  \BibitemOpen
  \bibfield  {author} {\bibinfo {author} {\bibfnamefont {E.}~\bibnamefont
  {Abe}}, \bibinfo {author} {\bibfnamefont {A.~M.}\ \bibnamefont {Tyryshkin}},
  \bibinfo {author} {\bibfnamefont {S.}~\bibnamefont {Tojo}}, \bibinfo {author}
  {\bibfnamefont {J.~J~L}\ \bibnamefont {Morton}}, \bibinfo {author}
  {\bibfnamefont {W.~M.}\ \bibnamefont {Witzel}}, \bibinfo {author}
  {\bibfnamefont {A.}~\bibnamefont {Fujimoto}}, \bibinfo {author}
  {\bibfnamefont {J.~W.}\ \bibnamefont {Ager}}, \bibinfo {author}
  {\bibfnamefont {E.~E.}\ \bibnamefont {Haller}}, \bibinfo {author}
  {\bibfnamefont {J.}~\bibnamefont {Isoya}}, \bibinfo {author} {\bibfnamefont
  {S.~A.}\ \bibnamefont {Lyon}}, \bibinfo {author} {\bibfnamefont {M.~L~W}\
  \bibnamefont {Thewalt}}, \ and\ \bibinfo {author} {\bibfnamefont {K.~M.}\
  \bibnamefont {Itoh}},\ }\bibfield  {title} {\enquote {\bibinfo {title}
  {{Electron spin coherence of phosphorus donors in silicon: Effect of
  environmental nuclei}},}\ }\href {\doibase 10.1103/PhysRevB.82.121201}
  {\bibfield  {journal} {\bibinfo  {journal} {Physical Review B - Condensed
  Matter and Materials Physics}\ }\textbf {\bibinfo {volume} {82}},\ \bibinfo
  {pages} {9--12} (\bibinfo {year} {2010})}\BibitemShut {NoStop}%
\bibitem [{\citenamefont {Witzel}\ \emph {et~al.}(2010)\citenamefont {Witzel},
  \citenamefont {Carroll}, \citenamefont {Morello}, \citenamefont
  {Cywi\'{n}ski},\ and\ \citenamefont {{Das Sarma}}}]{Witzel2010}%
  \BibitemOpen
  \bibfield  {author} {\bibinfo {author} {\bibfnamefont {Wayne~M.}\
  \bibnamefont {Witzel}}, \bibinfo {author} {\bibfnamefont {Malcolm~S.}\
  \bibnamefont {Carroll}}, \bibinfo {author} {\bibfnamefont {Andrea}\
  \bibnamefont {Morello}}, \bibinfo {author} {\bibfnamefont {{\L}ukasz}\
  \bibnamefont {Cywi\'{n}ski}}, \ and\ \bibinfo {author} {\bibfnamefont
  {S.}~\bibnamefont {{Das Sarma}}},\ }\bibfield  {title} {\enquote {\bibinfo
  {title} {Electron spin decoherence in isotope-enriched silicon},}\ }\href
  {\doibase 10.1103/PhysRevLett.105.187602} {\bibfield  {journal} {\bibinfo
  {journal} {Physical Review Letters}\ }\textbf {\bibinfo {volume} {105}},\
  \bibinfo {pages} {187602} (\bibinfo {year} {2010})}\BibitemShut {NoStop}%
\bibitem [{Note2()}]{Note2}%
  \BibitemOpen
  \bibinfo {note} {Another difference between the two $T_2$~measurements is the
  delay time between the end of the polarizing laser pulse and beginning of the
  Hahn echo sequence (\SI {900}{ms} for the echo-detected experiments and
  200--450~ms for the photoconductivity-detected ESR experiments). Charge
  reconfiguration and resulting Stark-field noise during the dark period could
  induce decoherence, however, while these effects have not been studied
  systematically, no difference in photoconductivity-measured $T_2$ has been
  observed for the two delay times \SI {200}{ms} and \SI
  {450}{ms}.}\BibitemShut {Stop}%
\bibitem [{\citenamefont {Dhar}\ and\ \citenamefont
  {Marshak}(1985)}]{Dhar1985}%
  \BibitemOpen
  \bibfield  {author} {\bibinfo {author} {\bibfnamefont {S.}~\bibnamefont
  {Dhar}}\ and\ \bibinfo {author} {\bibfnamefont {Alan~H.}\ \bibnamefont
  {Marshak}},\ }\bibfield  {title} {\enquote {\bibinfo {title} {{Static
  dielectric constant of heavily doped semiconductors}},}\ }\href {\doibase
  10.1016/0038-1101(85)90061-9} {\bibfield  {journal} {\bibinfo  {journal}
  {Solid-State Electronics}\ }\textbf {\bibinfo {volume} {28}},\ \bibinfo
  {pages} {763--766} (\bibinfo {year} {1985})}\BibitemShut {NoStop}%
\bibitem [{\citenamefont {Bube}(1992)}]{Bube1992}%
  \BibitemOpen
  \bibfield  {author} {\bibinfo {author} {\bibfnamefont {Richard~H}\
  \bibnamefont {Bube}},\ }\href@noop {} {\emph {\bibinfo {title}
  {Photoelectronic properties of semiconductors}}}\ (\bibinfo  {publisher}
  {Cambridge University Press},\ \bibinfo {year} {1992})\BibitemShut {NoStop}%
\end{thebibliography}%
